\pdfoutput=1
\documentclass[%
 reprint,
 amsmath,amssymb,
 aps,
]{revtex4-2}

\usepackage{dcolumn}
\usepackage{color}
\usepackage{amsmath}
\usepackage{xcolor}
\usepackage{amsfonts}
\usepackage{braket}
\usepackage{esint}
\usepackage{bm}
\usepackage{cancel}
\usepackage{hyperref}
\usepackage{graphicx}
\usepackage{caption}
\usepackage{subcaption}
\usepackage[resetlabels,labeled]{multibib}

\hypersetup{hypertex=true,
            colorlinks=true,
            linkcolor=blue,
            anchorcolor=blue,
            citecolor=blue}
\usepackage{enumerate}
\begin{document}

\preprint{APS/123-QED}

\title{Intrinsic instabilities in Fermi glasses}

\author{Yat Fan Lau}
\author{Tai Kai Ng}
\email{phtai@ust.hk}
\affiliation{%
Department of Physics, Hong Kong University of Science and Technology, Clear Water Bay Road, Kowloon, Hong Kong }%


\begin{abstract}
We study in this paper the effect of weak, short-ranged interaction on disordered metals. Through analysing the interaction matrix elements between different eigenstates of the non-interacting and corresponding Hartree-Fock single-particle Hamiltonian, we argue that as a result of localized single-particle eigenstates around the Fermi surface, the quasi-particle states on the Fermi surface are unstable towards formation of magnetic moments for arbitrary weak (but finite) repulsive interaction in the thermodynamic limit. This is a mechanism very different from the case of strong interaction $U\sim W_B$ ($W_B=$ bandwidth) or the quantum Griffiths effect where local moments are formed at small localized regions where coupling to the surrounding is weak. Numerical simulations are performed to verify our analysis. We further propose within a Landau Fermi-liquid-type framework that our result is applicable for general electronic systems with weak, short-ranged interaction as long as the quasi-particle states exist and are localized. An analogous result is obtained for attractive interaction, suggesting that Fermi glass state is intrinsically unstable in arbitrary dimension. 

\end{abstract}

\maketitle

\subsection{\label{sec:level1}Introduction} 
Interacting disordered systems have been a major areas of research in condensed matter physics because of the many interesting physics associated with disordered systems, including Fermi Glass\cite{anderson2004,freedman1977,mahmood2021,bhatt1984insulating}, Coulomb Glass\cite{efros1975coulomb,efros1976coulomb,ovadyahu2013interacting}, Wigner crystal\cite{falson2022competing,hossain2020observation}, the Many-Body Localized (MBL) states\cite{nandkishore2015many,abanin2019colloquium,abanin2017recent} and the anomalous metal states\cite{spivak2008theory,kapitulnik2019colloquium, wang2023}, etc. The problem is difficult theoretically because of the absence of an effective theoretical framework where interaction and disorder can be treated together systematically.  

Based on an analogy with Fermi liquid theory Anderson proposed that for weak enough short-ranged repulsive interactions, the ground and low energy excited states of disordered fermion systems with localized single-particle states evolve continuously as the interaction is turned on\cite{anderson2004,freedman1977} and the ground state remains a paramagnetic insulator with localized quasi-particles (Fermi glass) (see also Ref.\cite{basko2006metal}). 

However, using a linear stability analysis, Milovanovi{\'c} {\em et al.} found that in the presence of disorder, instability towards local magnetic moment formation occurs for strong enough repulsive interaction $U\sim W_B$ ($W_B=$ bandwidth) near metal-insulator transition or at small spatial regions weakly coupled to the surrounding (a quantum Griffiths effect)\cite{milovanovic1989effective}. More recently, Pilati and Fratini\cite{{pilati2016ferromagnetism}} found in quantum Monte Carlo calculations that the instability towards the formation of local moments occurs when the Fermi surface crosses the mobility edge. Experimentally, there are extensive studies on local moment formation in doped semiconductor materials such as Si:P etc.\cite{paalanen1986spin,alloul1987spin,ootuka1990static,hirsch1992esr}.

Other approaches to the problem include calculations on a cluster of a few (up to eight) spins by Walstedt \emph{et al.}\cite{walstedt1979magnetic}, renormalization group approach\cite{castellani1984spin,finkel1984weak} and quantum Monte Carlo method\cite{pilati2016ferromagnetism}.

In this paper, we point out that instability towards magnetic moment formation occurs rather naturally for single-particle states on the Fermi surface when the states are localized. In this case, magnetic instability occurs across states with energy $|\xi_{a}|\leq \bar{U}/L^d$ on Fermi surface for repulsive interaction, where $\bar{U}\sim$ strength of interaction, and $L^d$ is the localization volume in units of lattice spacing. This occurs even for weak interaction $\bar{U} \ll W_B$. An analogous result also applies for attractive interaction. The result is a direct consequence of the qualitatively different nature of interaction matrix elements between localized electronic states when compared with extended states and is not related to statistically rare events associated with quantum Griffiths effect. The formation of magnetic moment on Fermi surface can be demonstrated in Hartree-Fock (HF) or first-order perturbation theory.

To understand this surprising result, we study the effects of weak, short-ranged interaction on disordered electronic systems. In section \ref{IME} we shall analyze the interaction matrix elements between the eigenstates of the non-interacting and the corresponding Hartree-Fock single-particle Hamiltonians, and point out the qualitative difference between effects of interaction on localized and extended electronic states. In section \ref{EH} and \ref{instability} we construct the effective Hamiltonians in first-order perturbation and Hartree-Fock theory and show that the paramagnetic ground state formed by doubly occupying the lowest $2N$ effective single particle energy levels is unstable to spontaneous formation of local magnetic moments for repulsive interactions (and correspondingly, local fermion pairs for attractive interaction) across states with energy $|\xi_{a}|\leq \bar{U}/L^d$  on the Fermi surface as a result of the specific nature of the interaction matrix elements between localized states. This instability is not detectable by usual linear stability analysis\cite{milovanovic1989effective} because of an intrinsic limitation of linear response theory we explain in detail in Appendix \ref{AppB}. Our result is summarized in section \ref{LD} where we propose within a Fermi-liquid type framework that our result is expected to remain robust for general shot-ranged, repulsive interaction as long as quasi-particle states exist and are localized around the Fermi surface. 
\subsection{Interaction matrix elements}
\label{IME}
In general, the Hamiltonian of a disordered fermionic system can be written as $H=H_0+H'$ where $H_0$ is the non-interacting disordered Hamiltonian and $H'$ represents interaction between fermions.  
For concreteness, we consider a general lattice spin-$1/2$ fermion model of form
\begin{equation}
\begin{aligned}
    H_0 &= -\sum_{ij\sigma}t_{ij}(c^\dagger_{i\sigma}c_{j\sigma}+h.c.)
           +\sum_{i\sigma}(W_i-\mu) n_{i\sigma} \\
    H' &= \sum_{i\sigma j \sigma'}\frac{1}{2}U^{\sigma\sigma'}_{ij}n_{i\sigma}n_{j\sigma'} \label{eq1},
\end{aligned}
\end{equation}
where $c(c^\dagger)_{i\sigma}$ are spin-$\sigma$ fermion operators on lattice site $i$, $\sigma=\pm\frac{1}{2}$ and $\mu$ is the chemical potential. The first term in $H_0$ represents the hopping of fermions between different lattice sites with hoping matrix element $t_{ij}$ and the second term represents an onsite disordered potential $W_i$ which can be chosen to be a random variable distributed uniformly between $-W/2$ and $W/2$. $H'$ describes the interaction between fermions. We consider short-ranged interactions in this paper where $U^{\sigma\sigma'}_{ij}$ is nonzero only when the distance $|\vec{r}_i-\vec{r}_j|$ is less than a few lattice sites.

The (single-particle) eigenstates of the non-interacting Hamiltonian $H_0$ are given by 
\begin{equation}
    \sum_j(h_{0ij}-\mu\delta_{ij})\phi^a(\vec{r}_j) = \xi_{a}\phi^a(\vec{r}_i)
\end{equation}
where 
$h_{0ij}=-t_{ij}+W_i\delta_{ij}$. The Hamiltonian can be written in the eigenstate representation of the fermion operators $c(c^\dagger)_{a\sigma}$ defined by
\begin{equation}
    c_{i\sigma} = \sum_{a}\phi^a(\vec{r}_i) c_{a\sigma} \quad \text{and} \quad  c^{\dagger}_{i\sigma} = \sum_{a}\phi^{a*}(\vec{r}_i)c^{\dagger}_{a\sigma}.
\end{equation}
Substituting the above representations into Eq.(\ref{eq1}), we obtain the Hamiltonian in $H_0$-eigenstate representation,
\begin{equation}
    H = \sum_{a\sigma}\xi_a c^\dagger_{a\sigma} c_{a\sigma}+\frac{1}{2}\sum_{\substack{abce  \\ \sigma \sigma'}} U_{abce}^{\sigma \sigma'} c^\dagger_{a\sigma}c_{b\sigma}c^{\dagger}_{c\sigma'}c_{e\sigma'}
\end{equation} 
where 
\[ U_{abce}^{\sigma \sigma'} = \sum_{ij}U_{ij}^{\sigma \sigma'}\phi^{a*}(\vec{r}_i)\phi^b(\vec{r}_i)\phi^{c*}(\vec{r}_j)\phi^e(\vec{r}_j). \]
We emphasize that the indices $a,b,c,e$ are eigenstate indices, and are not related to momenta of particles.  \\ \\
For comparison, we also introduce the eigenstates of the HF Hamiltonian, given by
\begin{subequations}
\label{HHF}
\begin{equation}
\begin{aligned}
    H_{HF} &= -\sum_{ij\sigma}t^{HF}_{ij}(c^\dagger_{i\sigma}c_{j\sigma}+h.c.)
           +\sum_{i\sigma}W_i^{HF} n_{i\sigma}
\end{aligned}
\end{equation}
where
\begin{equation}
\begin{aligned}
    t_{ij}^{HF} &= t_{ij}+U_{ij}^{\sigma\sigma}\braket{c^\dagger_{j\sigma}c_{i\sigma}} \\
    W_i^{HF} &= W_i+\sum_{j\sigma'}U^{\sigma\sigma'}_{ij}\braket{n_{j\sigma'}},
\end{aligned}
\end{equation}
\end{subequations}
where $\braket{\cdots}$ denotes self-consistently determined ground state expectation value. We have assumed that $U_{ij}^{\uparrow\uparrow}=U_{ij}^{\downarrow\downarrow}$, $U_{ij}^{\uparrow\downarrow}=U_{ij}^{\downarrow\uparrow}$ and a non-magnetic HF ground state is found in the self-consistent HF theory, with $\braket{c^\dagger_{j\uparrow}c_{i\uparrow}}=\braket{c^\dagger_{j\downarrow}c_{i\downarrow}}$ and $\braket{n_{j\uparrow}}=\braket{n_{j\downarrow}}$. For attractive interaction, a Cooper instability occurs which we shall discussed later. The HF single-particle states are given by
\begin{equation}
\sum_j(h_{ij}^{HF}-\mu\delta_{ij})\phi^{\alpha}_{HF}(\vec{r}_j) = E_{\alpha}\phi^{\alpha}_{HF}(\vec{r}_i)
\end{equation}
where $h_{ij}^{HF}=-t_{ij}^{HF}+W_i^{HF}\delta_{ij}$. We introduce also the HF-eigenstate representation of the fermion operators $c(c^\dagger)_{\alpha\sigma}$, given by
\begin{equation}
    c_{i\sigma} = \sum_{\alpha}\phi^{\alpha}_{HF}(\vec{r}_i) c_{\alpha\sigma} \quad \text{and} \quad  c^{\dagger}_{i\sigma} = \sum_{\alpha}\phi_{HF}^{\alpha*}(\vec{r}_i)c^{\dagger}_{\alpha\sigma}.
\end{equation}
In the following, we shall analyze the interaction matrix elements $U^{\sigma \sigma'}_{abce}$ between eigenstates $\phi^a(\vec{r}_i)$'s of $H_0$ and $U^{\sigma \sigma'}_{\alpha\beta\gamma\zeta}$ between eigenstates $\phi^{\alpha}(\vec{r}_i)$'s of $H_{HF}$. We shall see that our analysis gives the same qualitative result in both cases as long as the single-particle states are localized. 

We first consider localized eigenstates $\phi^a(\vec{r}_i)$ of $H_0$, In this case we may write
\begin{equation}
    \phi^a(\vec{r}_i) \sim \frac{1}{L^{d/2}}e^{-\frac{|\vec{r}_i-\vec{x}_a|}{2L}} \quad \text{and} \quad |\phi^a(\vec{r}_i)|^2 \sim \frac{1}{L^{d}}e^{-\frac{|\vec{r}_i-\vec{x}_a|}{L}}
\end{equation}
for $|\vec{r}_i-\vec{x}_a| \gg L$ where $\vec{x}_a$ is the centre of the localized state $a$ and $L$ is the localization length. In this case, we expect there is small wavefunction overlap between states that are far from each other and the matrix element $U^{\sigma \sigma'}_{abce}$ decays exponentially as the distance between the states $|\vec{x}_p-\vec{x}_q| \gg L$ for any $(p,q)=(a,b,c,e)$. The matrix element is sizable only if $|\vec{x}_p-\vec{x}_q| < L$ for all $(p,q)=(a,b,c,e)$. In this case the order of magnitude of $U^{\sigma \sigma'}_{abce}$ can be estimated as
\begin{equation}
\begin{aligned}
    U^{\sigma \sigma'}_{abce}&=\sum_{ij}U_{ij}^{\sigma \sigma'}\phi^{a*}(\vec{r}_i)\phi^b(\vec{r}_i)\phi^{c*}(\vec{r}_j)\phi^e(\vec{r}_j) \\
    &\sim (\sum_{j}U^{\sigma \sigma'}(\vec{r}_i-\vec{r}_j))(\sum_{i}\phi^{a*}(\vec{r}_i)\phi^b(\vec{r}_i)\phi^{c*}(\vec{r}_i)\phi^e(\vec{r}_i)) \\ 
    &\sim \Bar{U}^{\sigma \sigma'}\frac{1}{L^{2d}}(L^d)=\frac{1}{L^d}\Bar{U}^{\sigma \sigma'} \label{Estimate U}
\end{aligned}
\end{equation}
where $\Bar{U}^{\sigma \sigma'} = \sum_{j}U^{\sigma \sigma'}(\vec{r}_i-\vec{r}_j)$ represents the ``average strength $\times$ the range of the interaction potential" and $\sum_{i}\phi^{a*}(\vec{r}_i)\phi^b(\vec{r}_i)\phi^{c*}(\vec{r}_i)\phi^e(\vec{r}_i)\sim L^d \sim$ volume where the wavefunctions overlap substantially. The estimation is valid as long as $L \gg$ range of the interaction potential. \\ \\
It is useful to look at the disorder-average of $U^{\sigma \sigma'}_{abce}$ which is given by 
\[ \braket{U^{\sigma \sigma'}_{abce}}_{dis} \sim {\Bar{U}^{\sigma \sigma'} \over L^d} \times P_{abce}, \]
where $P_{abce} \sim (\frac{L^d}{V})^3$ is the probability of finding all four states within distance $L$ from each other and $V$ is the volume of the system\cite{lee1985disordered}.

Notice that for $a=b$ and $c=e$, $U^{\sigma \sigma'}_{aacc} = \sum_{ij}U^{\sigma \sigma'}(\vec{r}_i-\vec{r}_j)|\phi^a(\vec{r}_i)|^2|\phi^c(\vec{r}_j)|^2\sim \frac{1}{L^d}\Bar{U}^{\sigma \sigma'}$ and $\braket{U^{\sigma \sigma'}_{aacc}}_{dis} \sim {\Bar{U}^{\sigma \sigma'} \over L^d} \times P_{aacc}\sim (\frac{\Bar{U}^{\sigma \sigma'}}{V})$ is much larger than $\braket{U^{\sigma \sigma'}_{abce}}_{dis}$ because the probability of finding two states within distance $L$ ($\sim L^d/V$) is much larger than the probability of finding four states within distance $L$; i.e., 
\begin{equation}
    \braket{U^{\sigma \sigma'}_{aacc}}_{dis} \sim (\frac{\Bar{U}^{\sigma \sigma'}}{V}) \quad \gg \quad \braket{U^{\sigma \sigma'}_{abce}}_{dis} \sim (\frac{\Bar{U}^{\sigma \sigma'}}{V})(\frac{L^d}{V})^2
    \label{dis_avg}.
\end{equation}
In particular, $\braket{U^{\sigma \sigma'}_{aaaa}}_{dis}\sim \frac{1}{L^d}\Bar{U}^{\sigma \sigma'}$ as there is no probability correction factor associated with finding the same state $a$. 

We now compare the above results with the case of extended states where
\begin{equation}
    \phi^a(\vec{r}_i) \propto \frac{1}{\sqrt{V}}e^{i u_a(\vec{r}_i)} \quad  \text{and} \quad |\phi^a(\vec{r}_i)|^2 \propto \frac{1}{V}
\end{equation}
where $u_a(\vec{r}_i)$ is a real number function and we obtain $U^{\sigma \sigma'}_{abce}\sim U^{\sigma \sigma'}_{aacc} \sim U^{\sigma \sigma'}_{aaaa}\sim \Bar{U}^{\sigma \sigma'}/V$.

Physically, a fermion in extended state $a$ interacts with all other fermion states in the system with matrix element of order $\Bar{U}/V$, and its properties are determined by the collective behaviour of the whole system. However, for localized states, a fermion in state $a$ interacts only with states within a distance $L$, but with a much stronger interaction $\sim \Bar{U}/L^d$. Its behavior is determined by local fermion environment in this case. 

It is quite obvious that our result is independent of the precise form of $H_0$ or the detailed behaviour of single-particle wavefunctions. It depends only on the assumption of localization of the single-particle wavefunction and the interaction potential is short-ranged. In particular, the same qualitative conclusion would be obtained for eigenstates $\phi^{\alpha}(\vec{r}_i)$ of $H_{HF}$ as long as the eigenstates of $H_{HF}$ are also localized. 

\begin{figure}
\centering
\includegraphics[width=\columnwidth]{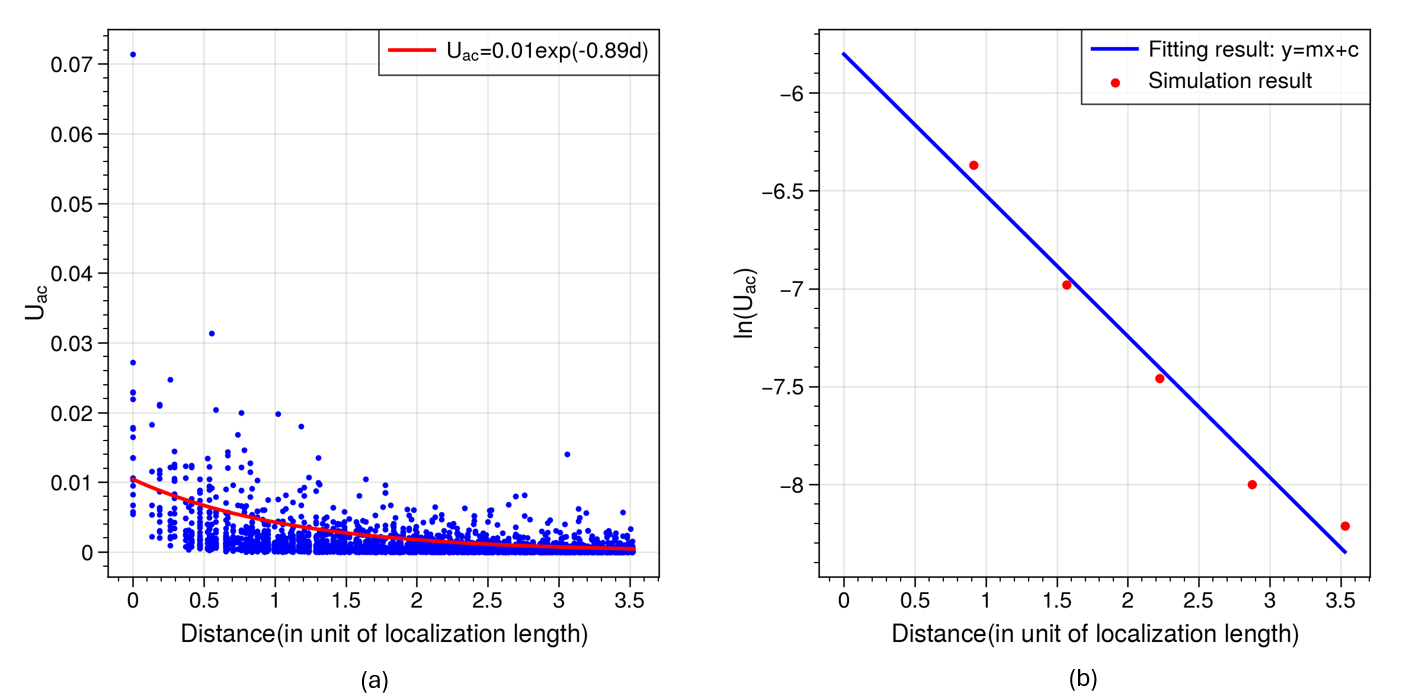}
\caption{(a)Simulation raw data for $U^{\sigma \sigma'}_{aacc}$ as a function of distance where $a$, $c$ are eigenstates of $H_0$. (b)The average $\langle U^{\sigma \sigma'}_{aacc} \rangle_{dis}$ as a function of distance}
\label{Vnm}
\end{figure}

To verify the above result, we perform a numerical simulation of $U^{\uparrow\downarrow}_{aacc}$ where $a$ and $c$ are eigenstates of $H_0$ on a $t$-$t'$-$W$-Hubbard-model on a $60 \times 60$ square lattice,
\begin{equation}
\begin{aligned}
    H =& -t\sum_{\langle ij \rangle \sigma}(c^\dagger_{i\sigma}c_{j\sigma}+h.c.)-t'\sum_{\langle \langle ij \rangle \rangle \sigma}(c^\dagger_{i\sigma}c_{j\sigma}+h.c.) \\
    &+\sum_{i\sigma} (W_i-\mu) n_{i\sigma} + U\sum_{i}n_{i\uparrow}n_{i\downarrow}
\end{aligned}
\label{t-t'-W-U}
\end{equation}
where the $t$, $t'$ terms are the nearest and next nearest neighbor hopping, respectively. We choose $t=1$, $t'=0.6$ , $\mu=0$ and $W=9$ in our simulation , corresponding to an average localization length $L\sim 7.65$\cite{wegner1980inverse}. The same set of parameters is used in all our numerical simulations below. The matrix element $U^{\uparrow\downarrow}_{aacc}$ is computed as a function of distance $d$ between two eigenstates $a$ and $c$ where $d$ is defined as the separation between max($|\phi^a(\vec{r}_i)|$) and max($|\phi^c(\vec{r}_i)|$). 
The results are shown in Fig.\ref{Vnm}(a) where each data point corresponds to a randomly chosen pair of states $(a,c)$ from 400 states near the band centre with energy range from $\sim -0.91t$ to $\sim 0.26t$ for five randomly chosen disorder configurations. The average $\langle U^{\uparrow\downarrow}_{aacc}\rangle$ as function of distance is shown in Fig.\ref{Vnm}(b) where we find that although there exists large fluctuations in $U^{\uparrow\downarrow}_{aacc}$, $\langle U^{\uparrow\downarrow}_{aacc}\rangle$ decays exponentially as expected with a decay factor $\sim 1.12L$ which is close to the localization length $L$.

We shall show in the next two sections that the volume {\em independence} of the average interaction matrix elements $\braket{U_{aaaa}^{\sigma\sigma'}}$ for localized states is what leads to the magnetic instability of states around Fermi surface. This instability does not occur for extended states.

\subsection{Effective Hamiltonian in First-order perturbation and HF theory}
\label{EH}
For very weak interaction we can study the system in first-order perturbation theory. In this case the wavefunctions of the system remain as eigenstates of $H_0$ and $H'$ only provides energy corrections to the eigenstates which are specified by the occupation numbers $\{n_{a\sigma}\}$ of $H_0$.  The resulting first-order energy takes the form
\begin{equation}
\label{eper}
\begin{aligned}
    E^{(1)} &= \braket{\Phi_{0}|H|\Phi_{0}} \\
    &=\sum_{a\sigma} \xi_an_{a\sigma} + \frac{1}{2}\sum_{ac\sigma \sigma'}(U^{\sigma \sigma'}_{aacc}-\delta_{\sigma \sigma'}U^{\sigma \sigma}_{acca})n_{a\sigma}n_{c\sigma'}. \\
\end{aligned}
\end{equation}

It is interesting to note that the same form of energy is obtained in the HF theory where the many-body states $|\Phi_{HF}\rangle$ are formed by occupying the single-particle eigenstates  of $H_{HF}$ ($\{\phi^{\alpha}_{\sigma}\}$) with energy 
\begin{equation}
\label{ehf}
\begin{aligned}
    E_{HF} &= \braket{\Phi_{HF}|H|\Phi_{HF}} \\
           &= \sum_{\alpha \sigma} \epsilon_{\alpha} n_{\alpha \sigma} + \frac{1}{2}\sum_{\alpha \beta \sigma \sigma'}(U^{\sigma \sigma'}_{\alpha \alpha \beta \beta}-\delta_{\sigma \sigma'}U^{\sigma \sigma}_{\alpha \beta  \beta \alpha })n_{\alpha \sigma}n_{\beta\sigma'}
      \end{aligned}
\end{equation}
where $n_{\alpha\sigma}$'s are the occupation number of the single-particle state $\phi^{\alpha}_\sigma$, $\epsilon_{\alpha}$ is the expectation value of $H_0$ for the single-particle state $\phi^{\alpha}_\sigma$ and $U^{\sigma \sigma'}_{\alpha \alpha \beta \beta} = \sum_{ij}U^{\sigma \sigma'}_{ij}|\phi^\alpha_{HF}(\vec{r}_i)|^2|\phi^\beta_{HF}(\vec{r}_j)|^2$\cite{ng2009introduction}. 

As $E^{(1)}$ and $E_{HF}$ has the same form of energy we shall simplify notation and use the same label $a$ and $c$ to denote single-particle eigenstates of both $H_0$ and $H_{HF}$ in the following when we consider the energies of different states in $E^{(1)}$ or $E_{HF}$.

An eigenstate of $E^{(1)}$ or $E_{HF}$ is characterized by a set of  occupation number $\{n_{a\sigma}\}$ of single-particle states. In both cases the energy required to add one particle to an unoccupied single-particle state $a$ is
\begin{subequations}
\begin{equation}
E_1(a)=\xi_a+\sum_{c\sigma'}(U^{\sigma \sigma'}_{aacc}-\delta_{\sigma \sigma'}U^{\sigma \sigma}_{acca})n_{c\sigma'}.
\end{equation}
whereas the energy for adding two particles in states $a$ and $c$ is
\begin{equation}
\label{e2}
E_2(a,c)=E_1(a)+E_1(c)+(U^{\sigma \sigma'}_{aacc}-\delta_{\sigma \sigma'}U^{\sigma \sigma}_{acca}). 
\end{equation}
Notice that an extra interaction term between the two added particles appears in $E_2(a,c)$.
\end{subequations}

A necessary condition for ground state stability is that $E_1(a)>0$ and $E_2(a,c)>0$ when we add particles to any (unoccupied) states $a,c$ in the ground state since otherwise we can lower the system's energy by adding one or two fermions to state $a, c$. Similarly, we expect $E_1(a)<0$ and $E_2(a,c)<0$ when we remove particles from any (occupied) states $a,c$ in the ground state.
We shall examine the stability of Fermi Glass state using these criteria in the following.

We note that as $H_{HF}$ is determined self-consistently from $\{\phi^{\alpha}_\sigma\}$, the values of $\xi_{\alpha}$, $U^{\sigma\sigma'}_{\alpha\alpha\beta\beta}$ and $U^{\sigma\sigma}_{\alpha\beta\beta\alpha}$ will be modified  slightly when we add or remove a few particles to/from the system. We shall see that these small differences will not affect our conclusion as long as the qualitative behaviour of the states $\{\phi^{\alpha}_\sigma\}$ are not modified by adding or removing a few particles to/from the system.
\subsection{Instabilities of the Fermi glass ground state}
\label{instability}
Naively the system's ground state $\ket{\Omega_0}$ is formed by doubly occupying the lowest $2N$ energy levels of $H_0$ (first order perturbation theory) or $H_{HF}$ (in HF theory). We shall show here that this state is unstable towards formation of local fermion pairs for attractive interaction in first order perturbation and in HF theory as long as there exist a finite density of states on the Fermi surface and the single-particle states we consider are localized. 

We start by adding a spin-$\sigma$ fermion to a state $a$ above the Fermi surface. We assume $E_1(a)>0$ as otherwise the Fermi glass state is already unstable. Now we add another electron with spin $-\sigma$ to the same state $a$. Using Eq.\ (\ref{e2}) the excitation energy in this case is 
\[
    E_2(a,a) = 2E_1(a) + U_{aaaa}^{\uparrow\downarrow}.
\]
 For attractive interaction, $U_{aaaa}^{\uparrow\downarrow}$ is negative and if $2E_1(a)<|U_{aaaa}^{\uparrow\downarrow}|$, we have $E_2<0$ which violates the stability criteria and the ground state becomes unstable. In this case the ground state of the system is formed by doubly occupying all states $a$ with energy $E_1(a)<|U_{aaaa}^{\uparrow\downarrow}|/2$. A single-particle excitation on the Fermi surface of the new ground state will acquire an energy gap $\sim |U_{aaaa}^{\uparrow\downarrow}|/2$ although the energy for adding a pair of $(\uparrow\downarrow)$ fermions on state $a$ remains gapless. 
 
Applying a similar analysis for repulsive interactions, it is easy to see that the Fermi glass ground state is unstable towards removing two fermions in a state $a$ below the Fermi surface with $E_1(a)<0$ if $E_2(a,a)=2E_1(a)+U_{aaaa}^{\uparrow\downarrow}>0$. In this case, we expect the real ground state is formed by doubly occupying all states $a$ with $E_1(a)<-U_{aaaa}^{\uparrow\downarrow}/2$ and singly occupying states $a$ with energy $-U_{aaaa}^{\uparrow\downarrow}/2<E_1(a)<0$, i.e., the ground state becomes {\em spin-polarized}. 

The above considerations have physical implication if there exists finite density of fermion states with $2|E_1(a)|<|U_{aaaa}^{\uparrow\downarrow}|\sim\frac{\Bar{U}^{\uparrow\downarrow}}{L^d}$ (see Eqs.(\ref{Estimate U}) and (\ref{dis_avg})). In this case the smallest plausible value of $|E_1(a)|$ is given by $|E_1(a)|\sim$ energy level spacing of the bulk system $\sim 1/(VN(0))$, indicating that such instability occurs generally when both $L$ and $N(0)$ are finite. Notice that the above analysis does not require large enough $\Bar{U}^{\uparrow\downarrow}$ or spatial regions with large fluctuations in the disorder strength.

The argument suggests that a finite portion of single-particle states with $|E_1(a)|<\frac{\Bar{U}^{\uparrow\downarrow}}{L^d}$ are unstable and the density of such states is of order $n\sim\frac{\bar{U}^{\uparrow\downarrow}}{2L^d}\times N(0)$. Notice that for extended systems $|U_{aaaa}^{\uparrow\downarrow}|\sim \frac{\Bar{U}^{\uparrow\downarrow}}{V}$ and the instability occurs only when $\Bar{U}^{\uparrow\downarrow}N(0)>2$, i.e., the instability occurs only for strong enough interaction. Therefore, we expect that for repulsive interaction a magnetic phase transition occurs when the Fermi surface crosses the mobility edge. For given $\Bar{U}^{\uparrow\downarrow}$ and $N(0)\neq0$. the density of magnetic moment is zero in the conducting phase where $L\rightarrow\infty$, and increases linearly with $L^{-d}$ in the insulating phase. 

We note that in reality spatial regions with large fluctuations in the disorder strength exists in disorder systems and magnetic moments associated with quantum Griffiths effect exists in general and would smear out the magnetic transition we discuss here. 

We can extend our analysis to instabilities associated with multiple particle excitation. As an example, we consider two particle excitation occupying two different states $a,c$.
In this case, following above argument, we expect that instability occurs if $U^{\sigma \sigma'}_{aacc} \sim U^{\sigma \sigma'}_{acca} \ge \bar{U}^{\uparrow\downarrow}/L^d$. However, this happens only when states $a$ and $c$ are within distance $L$. The probability for this to occur is of order $\sim \frac{U}{L^d}\times N(0)\times L^d$, and instability occurs only when $UN(0)\sim1$, which is much less important than the instability associated with excitation occupying the same state $a$. 

\begin{figure}[t]
\centering
\includegraphics[width=\columnwidth]{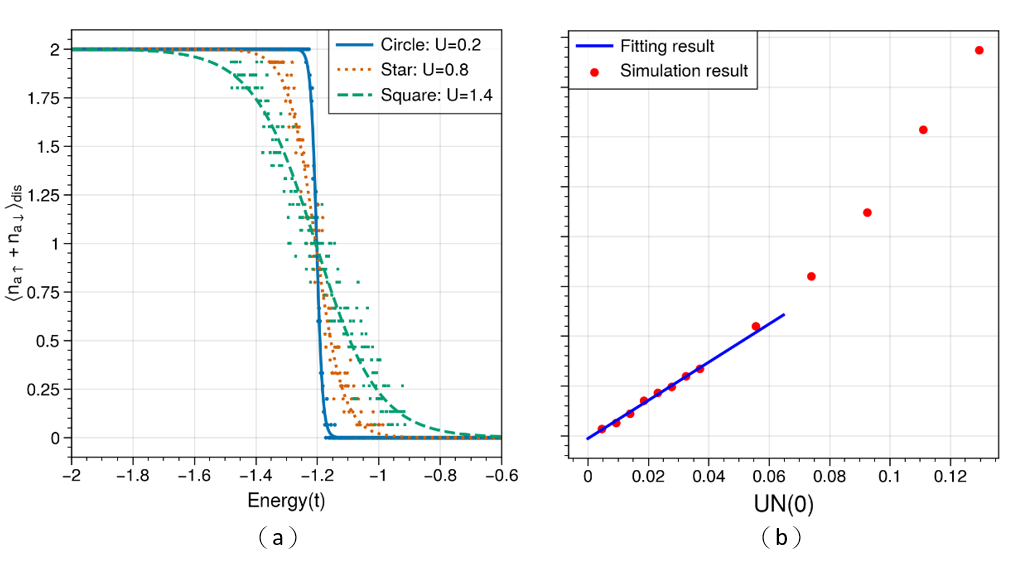}
\caption{(a)The average distribution $\braket{n_a}_{dis}=\braket{n_{a\uparrow}+n_{a\downarrow}}_{dis}$ over 15 disorder configurations for 3 different values of $U$, (b)The corresponding number of polarized spins as a function of $UN(0)$, where $N(0)$ is the density of states at Fermi surface}
\label{spin}
\end{figure}

To confirm our analysis we perform a Monte-Carlo simulation of the ground state of the repulsive $t$-$t'$-$W$-Hubbard-model (Eq.(\ref{t-t'-W-U})) in first-order perturbation theory where the ground state occupation numbers $\{n_{a\sigma}\}$ are determined by minimizing $E^{(1)}$ with respect to $\{n_{a\sigma}\}$. The results are shown in Fig.\ref{spin}(a) where we plot the average occupation numbers $\braket{n_a}_{dis}=\braket{n_{a\uparrow}+n_{a\downarrow}}_{dis}$ as a function of energy of the states $a$ and the data points are fitted by $\braket{n_a}_{dis}=\frac{2}{e^{\beta (\xi_a-\mu)}+1}$ with $\beta,\mu$ being fitting parameters (note that $\beta$ is not the temperature here). In this simulation, we choose total number of particles $\mathcal{N}=3000$ and perform $10^5$ Monte Carlo steps for each disorder configuration for 15 disorder configurations.  We obtain $\beta^{-1} \sim 2.63\frac{U}{L^2}$ for small $U$ in our simulation and the instability of the Fermi surface at energy range $|\xi_a|\leq U/L^d$ towards formation of local magnetic moments (in the weak interaction regime $UN(0)\ll 1$) is clear. We find that the number of polarized spins increases linearly with $U$ for small $U$. The number of polarized spins increases faster for larger $U$ (Fig.\ref{spin}(b)), indicating that instabilities associated with more than one single-particle states become important when $U$ increases. We emphasise here that this instability associated with the interaction matrix element $U_{aaaa}^{\uparrow\downarrow}$ is not detectable by usual linear stability analysis because of an intrinsic limitation of linear response theory which we shall explain in Appendix \ref{AppB}.

Our result is also supported by an earlier numerical work employing quantum Monte Carlo simulation on a continuous-space Hamiltonian for repulsive Fermi gas in which the system is subjected to a correlated speckle disorder\cite{pilati2016ferromagnetism}. It was shown that the system strongly favors the paramagnetic-to-ferromagnetic transition when the Fermi surface approaches the vicinity of the mobility edge. Our analysis suggests that besides the quantum Griffiths effect, the enhancement of the interaction diagonal matrix elements $U_{aaaa}^{\uparrow\downarrow}$  when quasi-particle states on Fermi surface become localized is another mechanism behind this observation. 

For attractive interaction, it is expected that a superconducting ground state will be formed for arbitrary weak interaction in a self-consistent BCS mean-field theory\cite{ma1985localized} and the effect of preformed local fermion (Cooper) pairs occupying the same state $a$ will be superseded. We show in a separate paper that the superconductor spectral gap is enhanced by these preformed local fermion pairs in a modified BCS theory\cite{2023replica}. 
\subsection{Beyond perturbation and Hartree-Fock analysis}
\label{LD}
The localization-induced instability of the Fermi glass state in first-order perturbation and HF theory lead us to consider the general situation of disordered spin-$1/2$ fermion systems with short-ranged repulsive interaction. Following a Fermi-liquid type analysis, Fleishman and Anderson \cite{fleishman1980interactions} argued that for weak, short-ranged interaction, the quasi-particle states on the Fermi surface can be described by an effective single-particle Hamiltonian with a generic form $H_{\text{eff}} = H_0 + \Sigma(\mu)$, where $\text{Re}\Sigma(\mu)$ is a random Hermitian matrix representing the interaction-induced self-energy on the Fermi surface and $\text{Im} \Sigma(\mu) = 0$ (stability of Fermi surface). The eigenstates of $H_{\text{eff}}$ represent the ``wave-function" of quasi-particles on the Fermi surface and are localized if $H_{\text{eff}}$ is ``random" enough in three dimensions (Fermi glass).

Assuming the validity of a Fermi-liquid type phenomenology we can write down a corresponding Landau energy functional describing low-energy excitations in the Fermi glass state, where
\begin{equation}
    \delta E_{\text{LD}}(\{\delta n_{a\sigma})\})=\sum_{a\sigma}\xi_a\delta n_{a\sigma}+{1\over2}\sum_{a\sigma;a'\sigma'}f_{a\sigma;a'\sigma'}\delta n_{a\sigma}\delta n_{a'\sigma'}
\label{landau}
\end{equation}
where $\delta n_{a\sigma}$ represents the fluctuation of quasi-particle occupation number $n_{a\sigma}$ on the Fermi surface, $\xi_a$ is the single quasi-particle energy and $f_{a\sigma;a'\sigma'}$ is the Landau interaction. Comparing Eq.\ (\ref{landau}) with Eq.\ (\ref{eper}) and Eq.\ (\ref{ehf}), we find that
$\xi_a=E_1(a)$ and
\[ f_{a\sigma;a'\sigma'}=U^{\sigma \sigma'}_{aaa'a'}-\delta_{\sigma \sigma'}U^{\sigma \sigma}_{aa'a'a} \]
in first-order perturbation or HF theory, with $a$ and $a'$ denoting the eigenstates of $H_0$ or $H_{HF}$.
In Landau Fermi liquid theory language, the instability of the Fermi glass state in first-order perturbation or HF theory is associated with the volume-independence of forward scattering amplitude $f_{a\sigma;a-\sigma}$ which is of order ${\Bar{U}\over L^d}$. 
More generally, it is well known that singular forward scattering term $f_{a\sigma;a\sigma'}$  which remains finite when system volume $V\rightarrow\infty$ leads to instability of the Fermi liquid state as long as the density of states on the Fermi surface $N(0)$ is finite\cite{hatsugai1992,baskaran1991,morimoto2016,ng2019,phillips2020exact} and our results from first order perturbation and HF theory are just another examples of this general phenomenon.

More generally, we expect that the singular forward scattering amplitude we observe in our analysis is a general property of localized quasi-particle states in systems with weak, short-ranged interaction. For localized quasi-particles $a$ and $a'$, $f_{a\sigma;a'\sigma'}$ should not depend on the system volume and can only depend on the distance $d$ between the two localized states when $d \gg L$ if interaction does not induce long-ranged effective interactions between localized quasi-particles. In particular, for $a=a'$, the only length scale which is available is the localization length and we anticipate that $f_{a\sigma;a-\sigma}$ can only scale with the size of the quasi-particle wavefunction $\sim 1/L^d$, and instability always occur if both $L$ and the density of states $N(0)$ are finite, as demonstrated in our analysis. 

Summarizing, based on a careful analysis of interaction matrix elements in first-order perturbation and Hartree-Fock theory, we show the existence of instability of Fermi surface towards formation of local magnetic moments for short-ranged repulsive interaction (and formation of local Fermion pair for attractive interaction) if the eigenstates on the Fermi surface are localized. The instability is driven by electron localization and is not associated with quantum Griffiths effect or strong electron-electron interaction. We further propose within a Landau Fermi-liquid type framework that our result is applicable for general electronic systems with weak, short-ranged interaction as long as the quasi-particle states exist and are localized, suggesting that Fermi glass states are intrinsically unstable.

We also note that we have assumed implicitly that $U_{aaaa}^{\sigma(-\sigma)}$'s are of the same sign as $a$ changes when we refer to ``attractive" or ``repulsive" interaction. This is the case for Hubbard interaction we discuss in this paper. For general form of interaction $U_{ij}^{\sigma\sigma'}$ this assumption may not be correct and both polarized spins and local fermion pairs may exist together in the ground state (see also \cite{sachdev1998magnetic}). In this case, the system becomes frustrated and spin glass or superconducting glass states may occur. We shall address these more exotic possibilities in future papers.

Finally we note that our analysis can be generalized to systems with spin-orbit coupling and the similar conclusion will be reached as long as inversion symmetry is not destroyed as the Kramer's degeneracy will be preserved(see for example\cite{girvin2019modern}).

\begin{acknowledgments}
The authors acknowledge helpful comment from Zvi Ovadyahu on our manuscript. The project is supported by a special funding support from the School of Science, the Hong Kong University of Science and Technology.
\end{acknowledgments}
\bibliographystyle{unsrt}
\bibliography{References}

\onecolumngrid 
\newpage

\appendix

\setcounter{subsection}{0} 
\setcounter{equation}{0} 
\setcounter{figure}{0} 

\section{Failure of linear instability analysis}
\label{AppB}
In this Appendix, we explain why commonly used instability analysis based on linear response theory fails to detect the Fermi Glass instability we discuss in this paper. We start with the Hartree-Fock(HF) Hamiltonian for the Hamiltonian Eq.(\ref{t-t'-W-U}) in the main text with repulsive Hubbard interaction $U$ 
\begin{equation}
    H_{HF} = -\sum_{ij,\sigma} t_{i j} c_{i \sigma}^{\dagger} c_{j \sigma}+\sum_{i,\sigma}\left(W_i+U \langle n_{i,-\sigma}\rangle\right) c_{i \sigma}^{\dagger} c_{i \sigma}
\end{equation}
Following Ref.\cite{milovanovic1989effective} we apply a small magnetic field $H_b=\sum_i B_i\left(n_{i \downarrow}-n_{i \uparrow}\right)$ to $H_{HF}$.

Following usual linear response theory, the change of an observable $A$ under perturbation is given by
$\delta\langle A(t)\rangle=\operatorname{Tr}(\delta \rho A)$, with $\delta \rho$ being the change in the density matrix due to $H_b$, where 
\begin{equation}
\label{lr}
    \delta \rho= \frac{i}{\hbar} \int_{-\infty}^t \mathrm{d} t^{\prime} \left[\rho_{HF} , H_{bI}(t')\right] 
\end{equation}
from Schr$\Ddot{o}$dinger equation and the corresponding change in magnetization is given by
\begin{equation}
\label{obs}
\langle \delta M(t) \rangle=\frac{i}{\hbar} \int_{-\infty}^t \mathrm{d} t^{\prime} \operatorname{Tr}\{\rho_{HF}\left[H_{bI}(t'),M_I(t)\right] \}
\end{equation}
where the subscript $I$ denotes the interaction picture 
in the eigenstate basis of $H_{HF}$. Introducing $c_{i \sigma}=\sum_a \phi^a\left(\vec{x}_i\right) c_{a \sigma}$, where $\phi^a\left(\vec{x}_i\right)$ are the single-particle eigenstates of $H_{HF}$, the magnetization operator $M_i$ and the perturbation $H_b$ can be written as
\begin{subequations}
\label{e4}
\begin{equation}
M_i=n_{i\uparrow}-n_{i\downarrow}
=\sum_{a a'}M_{aa'}\phi^{a*}\left(\vec{x}_i\right) \phi^{a'}\left(\vec{x}_i\right)
\end{equation}
and
\begin{equation}
H_b=-\sum_{aa'} B_{aa'} M_{aa'}
\end{equation}
\end{subequations}
where $M_{aa'}=c_{a \uparrow}^{\dagger} c_{a' \uparrow}-c_{a \downarrow}^{\dagger} c_{a' \downarrow}$ and $B_{aa'}=\sum_i B_i \phi^{a*}\left(\vec{x}_i\right) \phi^{a'}\left(\vec{x}_i\right)$. Substituting the above expressions into Eq.(\ref{obs}) and taking Fourier transform we obtain for time-independent $B_i$'s,
\begin{equation}
    \braket{\delta M_{aa'}} 
    = \sum_{aa'} \chi_{aa'}B_{aa'}
\end{equation}
 where 
\begin{equation}
\begin{aligned}
\label{sus}
    \chi_{aa'} &\sim -  \frac{f(E_a)-f(E_{a'})}{E_a-E_{a'}},
\end{aligned}
\end{equation}
$f(E_a)$ being the Fermi-Dirac distribution and $E_a$ is the HF single-particle energy of state $a$.
Replacing $B_i\rightarrow B_i+U\delta\braket{M_i}$ and correspondingly
$B_{aa'}\rightarrow B_{aa'}+\sum_{c,c^{\prime}}U_{aa^{\prime}cc^{\prime}}\braket{\delta M_{cc^{\prime}}}$
we obtain the self-consistent equation
\begin{equation}
\braket{\delta M_{aa^{\prime}}} 
    = \sum_{aa'} \chi_{aa'}\left(B_{aa^{\prime}}+\sum_{c,c^{\prime}}U_{aa^{\prime}cc^{\prime}}\braket{\delta M_{cc^{\prime}}}\right)
\end{equation}
which is the same result as that in Ref.\cite{milovanovic1989effective}, with the susceptibility matrix written in eigenstate basis. As $\chi_{aa}=0$ it is clear that the diagonal interaction matrix $U_{aaaa}$ does not contribute to the self-consistent equation. However, the $a=a'$ term is exactly what is responsible for the Fermi glass instability we discuss in this paper. 
In the following, we explain why $\chi_{aa}=0$ in linear response theory. 

We consider a general many body system with Hamiltonian $H$ and eigensates $\ket{A}$. In the eigenstate basis, the density matrix $\rho$ at equilibrium is given by $\rho=\sum_{A} e^{-\beta E_{A}} \ket{A}\bra{A}$.
In the presence of a perturbation there are two possible changes to the density matrix which can be formally written as  
\begin{equation}
\begin{aligned}
\label{drho}
\delta \rho_1 + \delta \rho_2 
=\sum_{A}\delta\left(e^{-\beta E_{A}}\right)|A\rangle\langle A|+\sum_{A} e^{-\beta E_{A}}(|\delta A\rangle\langle A|+| A\rangle\langle\delta A|)
\end{aligned}
\end{equation}
The first term $\delta \rho_1$ represents a change in the thermodynamics factor $e^{-\beta E_{A}}$ without changing the states $|A\rangle$ whereas the second term $\delta \rho_2$ comes from the change in the eigenstates under external perturbation (dynamical effect). Ordinary linear response theory  only accounts for the second term in Eq.(\ref{drho}), i.e., (see for example Ref\cite{ng2009introduction} )
\[ \sum_{A} e^{-\beta E_{A}}(|\delta A\rangle\langle A|+| A\rangle\langle\delta A|)=\frac{i}{\hbar} \int_{-\infty}^t \mathrm{d} t^{\prime} \left[\rho_{HF} , H_{bI}(t')\right] \]
and $\delta \rho_1$ is absent in linear response theory. This is fine when calculating linear responses of the system under external perturbation starting from the ``correct" ground state. However, problem may arise when we apply linear response theory to instability analysis of ground state as we shall illustrate in the following example.


We consider a single-$a$ Hamiltonian $H_0$ with only one single-particle state $a$ and two spin species,
\begin{equation}
    H_0=\sum_\sigma \xi_a n_{a\sigma} + U n_{a\uparrow} n_{a\downarrow}.
\end{equation}
 and consider a small perturbation $H_b=B_a (n_{a\downarrow}-n_{a\uparrow})$ on the system.
 
 Treating $H_b$ as perturbation it is easy to see that $\delta \rho_2=0$ in ordinary linear response theory for any density matrix of form
\[   \rho=\sum_A a_A|A\rangle \langle A|,  \]
where $|A\rangle$ are eigenstates of $H_0$ and $a_A$'s are arbitrary coefficients. This is because $\rho$ commutes with $H_b$ (or $H_b$ does not modify the eigenstates $|A\rangle$), independent of the values of $a_A$'s and we would conclude that the system is stable when we apply linear stability analysis to this system, even when $a_A$ does not correspond to the density matrix of the system in thermal equilibrium. Notice that on the other hand, $H_b$ changes the eigen-energies $E_a$ and thus the thermodynamics factor $e^{-\beta E_a}$

More explicitly, $H_0$ can be diagonalized easily with the 
eigensates \{$\ket{a\uparrow -a\downarrow},\ket{0},\ket{a\uparrow},\ket{-a\downarrow}$\} with corresponding energies \{$2\xi_a+U,0,\xi_a-B_a,\xi_a+B_a$\} and it is easy to see that for $\xi_a<0$, the doubly-occupied state 
\{$\ket{a\uparrow -a\downarrow}$\} has higher energy than the singly occupied (spin-polarized) states when $\xi_a+U>0$ at $B=0$. Usual instability analysis based on linear response theory cannot detect the instability of the doubly occupied state as the doubly occupied state is an eigenstate of the system which is not modified by $H_b$. The instability of the doubly occupied state as ground state of the system is a pure {\em thermodynamic} effect.

Returning to our problem of stability of Fermi glass state, we see that the external perturbation can be written as
$\delta b=\delta b_1 +\delta b_2$, where
\begin{eqnarray}
\delta b_1 & = & \sum_{a} B_{a a}\left(c_{a \downarrow}^{\dagger} c_{a\downarrow}-c_{a \uparrow}^{\dagger} c_{a\uparrow}\right)  \\ \nonumber
\delta b_2 & = & \sum_{a \neq a^{\prime}} B_{a a^{\prime}}\left(c_{a \downarrow}^{\dagger} c_{a^{\prime} \downarrow}-c_{a \uparrow}^{\dagger} c_{a^{\prime} \uparrow}\right)
\end{eqnarray}
where the first term commutes with $H_{HF}$ and produces no effect in linear-response theory. The second $a\neq a'$ term predicts instability towards local moments formation only at regions where the impurity sites is weakly coupled to the rest of the system or when interaction $U\sim W_B$ is strong enough (Ref.\cite{milovanovic1989effective}). What we discover in this paper is that the effect associated with the $B_{aa}$ term (or the $U_{aaaa} \sim \langle U\rangle\frac{1}{L^d}$ term) should be examined more carefully for localized eigenstates $\phi^{a}$. We compare the eigenstate energies of the Hartree-Fock (and first order perturbation) Hamiltonian carefully, and show that instability towards formation of local moment (for repulsive $U$) at the Fermi surface occurs for arbitrary small (but finite) $U$ in the thermodynamics limit $V\rightarrow\infty$ because of the $U_{aaaa}$ term. The linear stability analysis based on usual linear response theory cannot detect this instability because of the intrinsic limitation of linear response theory.



\section{Some details of our numerical calculations}
\begin{figure}[h]
\centering
    \includegraphics[scale=0.37]{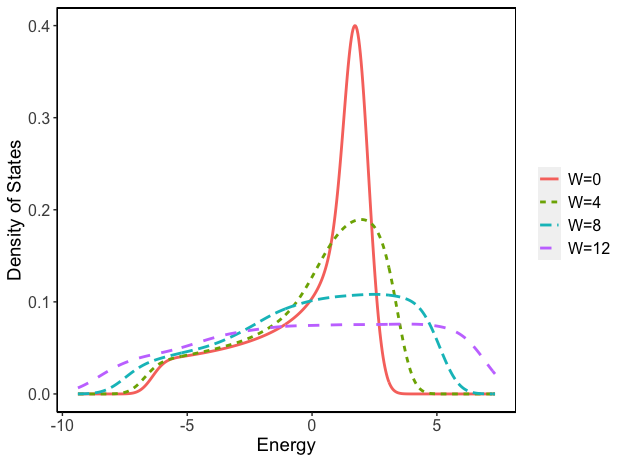}
    \caption{DOS per unit volume for different disorder strengths. We have set $t'=0.6t$ and taken average over 15 disorder configurations}
\label{DOS}
\end{figure}

In our simulation, we adopted a periodic boundary condition for the $(60 \times 60)$ square lattice. We set the disorder strength $W/t=9$ which is comparable to the bandwidth of the energy spectrum($\sim 8.8t$) at $W=0$. \\ \\
The density of states(DOS) per unit volume(area in 2D) with different disorder strengths for $t'=0.6t$ is shown in Fig. \ref{DOS}. The large peak in the DOS reflects the existence of Van Hove's singularity in the $W=0$ limit. 
\end{document}